\documentclass[aip,preprint,nofootinbib]{revtex4-1}
\usepackage{bm,natbib,amsmath,amssymb,wasysym,pifont}
\usepackage{tikz-cd}
\usepackage{xcolor}
\usepackage{amsbsy}
\usepackage{bbm}
\usepackage{esvect}
\usepackage{todonotes}
\usepackage{graphicx}
\definecolor{Blue}{rgb}{0,191,200}

\newcommand{\ldot}{\dot{\cal L}}
\newcommand{\bek}{\begin{equation*}}
\newcommand{\eek}{\end{equation*}}
\newcommand{\w}{\wedge}
 \begin{document}
 \newcommand{\bea}{\begin{eqnarray}}
\newcommand{\eea}{\end{eqnarray}}
 \title{Fluid dynamics in the spirit of Cartan:  a coordinate-free formulation for an inviscid  fluid in 
 inertial and non-inertial frames.}
 \author{Alberto Scotti}
 \affiliation{Dept. of Marine Sciences,
 University of North Carolina,
 Chapel Hill, NC 27599-3300}
 \date{\today}
\begin{abstract}
Using Cartan's  exterior calculus, we derive a coordinate-free formulation of the Euler equations. These equations are invariant under Galileian transformations, which constitute a global symmetry. 
With the introduction of an appropriate generalized Coriolis force, these equations become symmetric under general coordinate transformations. 

We show how exterior calculus simplifies dramatically the derivation of conservation laws. We also discuss the advantage of an exterior calculus formulation with respect to symmetry-preserving discretizations of the equations. 

\end{abstract}
\maketitle
\section{Introduction}
It may seem odd, if not presumptuous,  to write a paper about the Euler equations in 2016. A quantitative theory of Newtonian fluids was given by Stokes in 1845 \citep{Stokes45}, which gave us a way to implement Leibniz {\em calculemus} dictum to fluids.
The acceptance of vector calculus in the late 19th century resulted 
in the formulation found in countless publications, 
which is simply a rewriting of Stokes' original equations with a more compact notation, though  we still  often see in publications and lectures
the equations {\em just as Stokes wrote them} close to two centuries ago. 
It is certainly a testament to the genius of Stokes (and Heaviside and Gibbs, who introduced vector calculus)  that his equations 
have stood the test of time.

There is another reason, however, why the Navier-Stokes equations stand out.  In the history of physical sciences, the norm is that the language used to formulate theories  evolves.
One the one hand, this makes harder to access the original literature without a proper training. This is why it is very hard, for example, 
to read Galileo or Newton, since our calculus-imbued mental processes have a hard time adjusting to the geometrical, proportion-based exposition typical of these authors\footnote{Newton is credited to invent calculus, 
but it took nearly a century and a half and the development of analysis to arrive at the logically consistent notion of limit we enjoy today.}. 
On the other hand, it is only through the continuous interplay of the physical concepts and the mathematics that we use to describe them that both evolve.   In this respect,
the Navier-Stokes equations stands as the exception. 

From a theoretical point of view, vector calculus suffers several drawbacks, some of which can be ameliorate by switching to tensor calculus, of which vector calculus is a subset. However, 
tensor calculus cannot avoid seeing physically relevant quantities through the lens of a particular sets of coordinates. In other words, the primal idea that the laws of physics should
not depend on the specific choice of coordinates that we use to map space (e.g., Cartesian vs. spherical) is obfuscated by the tensor approach.

Starting with the work of H. Poincar\'e, \'E. Goursat and especially E. Cartan,
there has been a push to recover a coordinate-free formulation of differential geometry, which stands at the foundation of any field theory, using {\em exterior calculus}.  Physical laws formulated  with the language of exterior calculus express ideas on a level abstracted from any particular coordinate system that an observer may use to describe her space. This frees us to explore the properties that derives from the laws {\em qua} laws (e.g., conserved quantities, symmetries,\ldots), as opposed to particular properties that emerge only when the laws, expressed in a suitable coordinate system, are applied to a {\em sui generis} problem.   

The impetus for arriving at an exterior calculus formulation of fluid dynamics is not purely theoretical. Over the last 20 years  a discrete counterpart of exterior calculus, the so-called discrete geometry has emerged \cite{Desbrun05,JPO}. The goal is to have discrete objects that satisfy as many (if not all) of the same algebraic properties of their continuum counterparts. This offers the tantalizing prospect of being able to discretize conservation laws in a way that respects the underlying symmetries {\em regardless} of the geometry of the manifold.  

The use exterior calculus also brings a logical clarity that vector calculus lacks. For example, consider an intensive quantity such as temperature, and an extensive quantity such as the internal energy. In vector calculus, we represent both with the same object, a scalar field.  The objects of exterior calculus, called $p-$forms, where $p$ ranges from zero the the dimension of the manifold, belong to a graded algebra. As we shall see, intensive quantities are handled with $0-$forms, whereas intensive quantities are handled by $n-$forms. Likewise, velocity and vorticity are handled by vector calculus with the same object, a vector\footnote{Technically, vorticity is a pseudo-vector, but we
assume that the manifold is orientable, and we select an orientation {\em ab initio}, thus rendering the distinction moot.}, whereas in exterior calculus they are handled with $1-$ and $2-$forms respectively. Further, the use of exterior calculus eliminates a rarely mentioned, by nonetheless real, logical inconsistency in the standard derivation. When formulating the conservation of momentum in the standard approach, we arrive at  integrals over surfaces and volumes of vector quantities. But such integrals make no sense on general manifolds. For example, using spherical coordinates, how do we make sense of \textit{adding} vectors defined over an extended region?\footnote{Tensor calculus tells us how to calculate \textit{differences} between nearby vectors.} Normally, we retreat to Cartesian coordinates, but that leaves open the question of validity of the resulting equations in general coordinates. 
On the other hand, there is not problem with forms, because exterior calculus is  the theory that formalizes the notion of integrals over $p-$dimensional surfaces. 

It is possible to find "translations" of the Navier-Stokes equations into exterior calculus language \citep[see, e.g. ref. ][p. 484--487]{Westenholz81}, but they are formulated with \textit{vector}-valued forms. 
The goal of this paper, which is more pedagogical, is to provide a derivation, as opposed to a translation, of the Euler equations entirely  contained within the landscape of the exterior calculus of \textit{scalar}-valued forms.

The rest of the paper is organized as follows: In Section 2, we begin by introducing  the Cartan derivative, which specializes the Lie derivative\footnote{Which, in turns, extends the concept of material derivative to forms} to the appropriate subspace of $p-$forms in classical space-time. 
We then look at conservation laws for intensive and extensive quantities, how they relate to each other, which lead to the equation for mass conservation. We then switch to the structures ($1-$ and $2-$forms) that encode kinematic quantities (circulation and vorticity). Using an axiomatic approach based on Kelvin's theorem, we derive equations for the vorticity and circulation that express the dynamics.  At this point
we are in a position to derive equations for energy, potential vorticity and helicity. 
Abstracted from the spatial coordinates, it will become 
clear how the law of motion of fluid still depend on the relative motion of different observers (now seen as entities in space-time), leading to the introduction of a generalized Coriolis force that applies 
to arbitrary frames. The latter, as far as we know,  is a novel result presented in section 3, followed by a brief summary. 
For the readers who may not be familiar with exterior calculus and differential forms, an Appendix  provides a quick survey of the  concepts, operators and notation of exterior calculus.

 \section{Euler equations with exterior calculus}
 While it is certainly possible to "translate" the fluid dynamics equations
from their usual vector-calculus based formulation to a formulation based on exterior calculus,
we prefer to derive a theory of fluid dynamics using only the tools of exterior calculus.  A complete theory 
requires the use of \textit{vector}-valued differential forms, but this greatly increases the technical 
background. For this reason, in this paper  
we will restrict to systems that can be described solely with \textit{scalar}-valued differential forms. Also, we neglect diabatic processes (e.g., friction and diffusive processes). Since the focus is on the basic kinematic structures, this is not an overly restrictive simplification. 

\subsection{A few conventions regarding notation}
In the following, we will assume that the reader has a working knowledge of exterior calculus and the theory of Lie groups. Readers who may not have had previous exposure to exterior calculus should consider reading Appendix A and B first, which provide a quick summary to the relevant concepts and notation. 

While physical applications of the theory are restricted to 2- and 3-dimensional spaces,  one of the (many) advantages of exterior calculus is that the algebraic properties of the operators do not depend on the dimension of the underlying manifold. Therefore, there is no real advantage in specifying at the onset the dimensionality of the physical space. 
Thus, in the following, we adopt the convention of using $n$ to denote the number of spatial dimensions. 
A Greek lowercase letter with a superscript $p$ will indicate a generic $p-$form.   Lowercase Greek letters without a superscript will denote
1-forms, Greek uppercase letters will denote 2-forms, while Fraktur letters will be used 
for $n-$forms. Finally, scalars (0-forms) will be
denoted with Latin letters. There are a few exceptions to this rule: The letter $d$ is of course reserved for the exterior derivative;  $\rho$ has long been used to denote the scalar density field and we do not depart from that convention.
The fluid occupies a spatial manifold which will be assumed to be oriented, endowed with a metric structure, and whose volume element is
$\mathfrak{V}$. 

\subsection{Classical space-time: the Cartan derivative}
A $p-$form $\alpha^p$ associates to a $p-$surface belonging to a generic $n-$dimensional  manifold $\mathbb{M}$ (with $n\geq p$) a value. Consider a 1-parameter Lie group of transformations of $\mathbb{M}$ onto itself, described by its operator ${\bf v}\equiv v^i\bm{\partial/}\bm{\partial x}^i$ (in general, boldface notation will denote vector, understood as linear functionals on the space of $1-$forms). The Lie derivative ${\cal L}_{\bf v}(\alpha^p)$ measures the rate of change of $\int\alpha^p$ under the action of the 1-parameter Lie group whose operator is ${\mathbf{v}}$.  Here, we consider how this notion specializes
to the case where the manifold describes classical space-time. 

In classical space-time time plays a special, absolute role:  the time lapsed between two events is {\em the same} for all possible observers, regardless of their relative state. We can also 
introduce a distance between events provided both events occur  {\em at the same time}. However, it is meaningless to ask what is the distance between two events which occur at different times. 
Thus, the defining characteristic of classical space-time is that if two events are simultaneous  for one observer, they are simultaneous for {\em all} observers. Put it succinctly, we speak of absolute
simultaneity. 
 In order to conduct experiments, a generic observer associates to each point $P$ of space-time a set of spatial coordinates
 (observer will be used a shorthand for a set of coordinates). That is, the observer is endowed with one or more charts that locally assign a correspondence 
 between points of the manifolds $\mathbb{M}$ to points in one or more subsets of $\mathbb{R}^n$.
 The charts are completely arbitrary, and they can depend on time, in the sense that what one observer perceives as the same point in that its coordinates do not change with time, another observer may not, and are subject only to the standard requirement of continuity. 
 Time, on the other hand is, aside from a shift, the same for all observers. 
 For such an observer, classical space-time is $\mathbb{G}=\mathbb{R}\times \mathbb{M}$, and we call this a classical observer, or just an observer. 
 Of the $(n+1)$ coordinates,  $x^0,x^1,\ldots,x^n$, let  $x^1,\ldots,x^n$ be the spatial coordinates and $x^0$ the time coordinate, which, when needed, we  relabel $t$ to highlight its special role.  By convention Greek indexes will run from $0$ to $n$, whereas Latin indexes will span only the spatial coordinates. 
If $x^\alpha$'s are the coordinates of a point $P$ for one observer, the coordinates of the same point $\hat{x}^\alpha$ for a different observer are given by 
\begin{equation}
\hat{x}^\alpha=F^\alpha(x^0,\ldots,x^n).
\end{equation}
The absolute nature of simultaneity means that the $F^i$'s  can be arbitrary functions of the spatial coordinates and time (e.g., a Galileian boost), whereas $F^0(x^0,\ldots,x^n)=x^0+s$, where $s$ is an arbitrary constant.  
 We call  the transformations of space-time that maintain absolute simultaneity  {\em Ersatz} Gauge Transformations (EGTs)\footnote{Ersatz in the sense that they are not on the same plane as the Gauge transformations that grown up physicists use.}. Of particular interest are sets of EGTs that are Lie groups:   The standard 10-parameter Galileo group provides such an example;
another example, important for the discussion to follow,  is   the one-parameter Lie group whose operator is 
\begin{equation}
{\bf v}\equiv \frac{\bm{\partial}}{\bm{\partial t}}\unboldmath +v^i\bm{\frac{\partial}{\partial x^i}}
\end{equation}
where the $v^i$'s are as smooth as necessary functions of space-time. We call this a \textit{flow}. Another way to look at a flow is
as the vector field in space-time tangent to the trajectory of particles carried by the flow, measured by an observer who parameterizes  the trajectory with the time coordinate, that is 
\begin{equation}
1=v^0=\frac{dt}{ds},\,v^i=\frac{dx^i}{ds}.
\end{equation}


Since the goal is to build a physical theory of fluids based on forms, we now consider some of the implications of the specific nature of classical space-time to forms. 
We denote with 
\begin{equation}
G\equiv\overset{n+1}{\underset{p=0}{\oplus}}\bigwedge^p\mathbb{G}
\end{equation}
the graded algebra of forms in  classical space-time. 
For an observer,  we define the 1-parameter family of inclusion maps  
\begin{equation}
i_s: \mathbb{M}\to \mathbb{R}\times\mathbb{M}; (x^1,\ldots,x^n) \to (s,x^1,\ldots,x^n)
\end{equation}
where $s\in\mathbb{R}$ is the parameter. We leave to the reader  to prove that the set of forms 
\begin{equation}
T=\{\alpha^p\in G: \forall s, i_s^*\alpha^p=0\}
\end{equation} is an ideal of $G$.\footnote{An ideal of an algebra is a subalgebra of elements of the algebra which is closed relative to addition and scalar multiplication, and such that the product of an element of the ideal with an element of the algebra belongs to the ideal.  } Here $i_s^*$ is the pullback induced by the inclusion map.  
Furthermore, it is not difficult to realize that the ideal $T$ is closed under the action of an EGT (this is because under an EGT $dt\to dt$), and so all classical observers agree on $T$, meaning that if $\alpha^p\in T$ for one observer, its pullback to the coordinates of another observer belongs to $T$ as well. 

Next, we consider the quotient ring of G relative to  $T$ which we denote with $G/T$.\footnote{An ideal $T$ can be used to introduce a notion of congruence. So two forms $\alpha$ and $\beta$ are congruent if $\alpha-\beta\in T$, and one can say that $\alpha$ and $\beta$ belong to the same equivalence class. The union of all such equivalence classes is the quotient ring of the algebra relative to $T$. } The main assumption is that \textit{physical quantities are encoded by forms that belong to} $G/T$: Because simultaneity is absolute, {\em all observers can compare measurements} (here, intended as values of forms) {\em taken at the same time}. Table~\ref{tab:1} lists the different types of physical quantities and the corresponding forms in $G/T$ (Note that $\bigwedge^{n+1}\mathbb{G}\,\mbox{modulo}\, T =\emptyset$, thus we consider only forms of grade no greater than $n$).  
\begin{table}
\begin{tabular}{c |c |c}
\hline
Type & Example & Form\\
\hline
\hline
 intensive quantities & temperature, entropy, potentials,\ldots & 0\\
line integrals & forces, action,\ldots & 1\\
angular velocities & vorticity & 2\\
fluxes & mass flux, energy flux,\ldots & $n-1$\\
extensive quantities & energy density, mass density, volume,\ldots & $n$\\
\hline
\end{tabular}
\caption{\label{tab:1} Physical quantities and the forms that encode them.}
\end{table}

 We are going to consider how forms in $G/T$  behave under the action of a flow as defined above. 
We introduce the special notation 
\begin{equation}
\dot{d}\equiv\left(dt\frac{\partial}{\partial t}+dx^i\frac{\partial}{\partial x^i}\right),
\end{equation}
while 
\begin{equation}
d\equiv\left(dx^i\frac{\partial}{\partial x^i}\right).
\end{equation}
Note that on the $(n+1)-$dimensional manifold $\mathbb{G}$, $\dot{d}$ is the standard exterior derivative.  
We state the main result as two theorems, which are left to the reader to prove:\\
\textbf{Theorem I}  \\
\textit{
Let
\begin{equation}
\dot{\bf v}=\left(\bm{\frac{\partial}{\partial t}}+v^i\bm{\frac{\partial}{\partial x^i}}\right)
\end{equation}
be a flow in classical space-time. 
For a generic form $\alpha^p\in\mathbb{G}$ the Lie derivative 
\begin{equation}
\dot{d}i_{\dot{\bf v}}(\alpha^p\w dt)+i_{\dot{\bf v}}\dot{d}(\alpha^p\w dt)=(-1)^p\frac{\partial\alpha^p}{\partial t}\w dt+d(i_{\bf v}\alpha^p)\w dt.
\end{equation}
From this, it follows that the ideal $T$ is closed relative to $i_{\dot{\bf v}}\dot{d}+\dot{d}i_{\dot{\bf v}}$, that is  $i_{\dot{\bf v}}\dot{d}(T)+\dot{d}i_{\dot{\bf v}}(T)\subseteq  T$. } \\  
 Hint:  a generic non-zero element of $T$ is the exterior product of a member of the quotient ring with $dt$. \\
\textbf{Theorem II} \\
\textit{
If $\alpha^p\in G/T$ and $\dot{{\bf v}}$ is a flow in classical space-time, then   
\begin{equation}
i_{\dot{\bf v}}\dot{d}\alpha^p+\dot{d}i_{\dot{\bf v}}\alpha^p=\frac{\partial\alpha^p}{\partial t}+i_{\bf v}d\alpha^p+di_{\bf v}\alpha^p\,\,\mathrm{modulo}\,\,T,
\end{equation}
where 
\begin{equation}
{\bf v}= v^i\bm{\frac{\partial}{\partial x^i}}
\end{equation}
}
(the partial derivative $\partial /\partial t$ means that the coefficients are to be derived in time, but no $dt$'s appear). 

Thus, for forms in the quotient ring,  the action of a flow modulo $T$ is given by the operator
\begin{equation}
\dot{\cal L}_{\bf v}\equiv \frac{\partial}{\partial t}+{\cal L}_{\bf v},
\label{eq:Cartan}
\end{equation}
where ${\cal L}_{\bf v}$ is the Lie derivative acting on the spatial coordinates alone. We call $\ldot_{\bf v}$ the Cartan derivative. The Cartan derivative satisfies the Leibniz rule.  However, it is {\em not} linear in ${\bf v}$, unless the form is constant in time. It does commute with  $d$ (recall that $d$ is restricted to $x^1,\ldots,x^n$), but it does not commute with $i_{\bf v}$ unless both the form and the flow are $t$ independent. 

{\em To summarize, in a physical theory that encodes information in forms that belong to $G/T$, that is, all physically relevant information is "read" from forms via an inclusion pullback $i_s^*$ at a given time $s$, then the Cartan derivative is the operator that gives the rate of change of physical forms under a flow.} 

 When applied to a $0-$form (i.e., a scalar field), the Cartan derivative coincides with the standard material (Lagrangian) derivative of fluid mechanics.  It extends the material derivative concept to forms of higher grade, just as the exterior derivative $d$ subsumes the operators gradient, divergence and curl into a unified operator.

\subsection{Lagrangian Invariants and Conserved Currents}
Intensive quantities (e.g., temperature, enthropy,...) are described by $0-$forms. {\em A Lagrangian Invariant (LI) of the flow ${\bf v}$ is any $0-$form $a$ that satisfies}
\begin{equation}
\dot{\cal L}_{\bf v}(a)=0
\end{equation}
{\em everywhere on the manifold.}
As the name suggests, a LI is an intensive quantity that remains unchanged when measured by an instrument that moves with the flow.\\
Extensive quantities are by definition properties of a volume of matter that can be added (e.g.,  kinetic energy). We describe them with $n-$forms. \\
{\em A Conserved Current (CC) of the flow ${\bf v}$ is a pair of $n-$forms $(\mathfrak{A},\mathfrak{B})$ which satisfy }
\begin{equation}
\frac{\partial}{\partial t}(\mathfrak{A})=-{\cal L}_{\bf v}(\mathfrak{B})
\end{equation}
{\em everywhere on the manifold.}
A CC is the volumetric density of a property A such that the only way to change the amount of $A$ within a {\em fixed} volume is by material fluxes of the property B in and out of the boundary of the volume. A special case is when the second element of the pair is equal to the first. In such case, we call $\mathfrak{A}$ a Materially Conserved Current (MCC). 

We see here an example of a more general situation.  Intensive and extensive quantities are conceptually distinct entities. In the standard vector calculus approach, they are both described by scalar fields. With exterior calculus, they are described by different spaces of the graded algebra, $0-$forms for intensive quantities, $n-$forms for extensive quantities.

We leave to the reader to verify the following propositions:\\
\textbf{Proposition I}  Let ${\mathfrak{A}}$ be a MCC whose scalar concentration is $A=\star\mathfrak{A}$. Then 
\begin{equation}
\frac{\mathrm{d}}{\mathrm{d}t}\int_V (A\mathfrak{V})=- \int_{\partial V}(A\Phi),
\end{equation}
where the form $\Phi\equiv i_{\bf v}\mathfrak{V}$ and $V$ is a volume fixed in space. Hence, the corresponding flux across  
across an oriented surface $S$ is $A\Phi$.\footnote{Technically, fluxes are pseudo $(n-1)-$forms, unlike for example the vorticity, which is always a $2-$form. In practice, once we choose an orientation for the manifold (provided it is orientable!), pseudo forms and forms 
cannot be distinguished. } \\
\textbf{Proposition II} If the volume $n-$form is an MCC, the Hodge star of a LI is a MCC and vice-versa. \\
\textbf{Proposition III}
Let  $\mathfrak{B}$ be an MCC, and $(\mathfrak{A},\mathfrak{J+A})$ a CC. Let $r=\star\mathfrak{A}/(\star\mathfrak{B})$. Then 
\begin{equation}
\ldot_{\bf v}(r)=\frac{1}{(\star\mathfrak{B})}[\star{\cal L}_{\bf v}(\mathfrak{J})].
\end{equation}
It follows that  if ${\mathfrak{J}=0}$, so that ${\mathfrak{A}}$ is a MCC, then  the ratio of the Hodge star of two MCCs is a LI. 
 \subsection{Mass conservation}
The prototypical example of a MCC in fluid dynamics 
is the mass $n-$form $\mathfrak{M}$, which associates to a volume the mass contained in the volume. In classical physics mass is strictly conserved. Imagine that  at a certain arbitrary time $t_0$ an observer tags a certain volume, e.g. by assigning a certain value to a LI (e.g., a die).  Due to the absolute nature of simultaneity, all observers can agree on what has been tagged, even though different observers will described the evolution of the region differently.
Conservation of mass under a physically admissible flow ${\bf v}$ requires 
\begin{equation}
{\dot{\cal L}_{\bf v}(\mathfrak{M})=0.}
\label{eq:CAI}
\end{equation}
It is worth to recall that the Cartan  derivative measures the rate of change of a $p-$form integrated over an arbitrary $p-$surface as both evolve under the flow. Therefore the meaning of (\ref{eq:CAI}) is  that the total mass contained in an arbitrary volume does not change even as the volume is distorted by the flow.
To the mass form we can associate its Hodge dual  $\rho=\star\mathfrak{M}$, the scalar density.  
Note that in general 
\begin{equation}
\dot{\cal L}_{\bf v}(\rho)\neq0.
\end{equation}
If and only if (see Proposion II of previous section)
\begin{equation}
\dot{\cal L}_{\bf v}(\mathfrak{V})=0,
\label{eq:inc1}
\end{equation}
then 
\begin{equation}
\dot{\cal L}_{\bf v}(\rho)=0.
\label{eq:inc2}
\end{equation}

Equations (\ref{eq:inc1}) or (\ref{eq:inc2})  above shows how the incompressibility condition  is written in the language of forms\footnote{It may be worthwhile to point out the difference between an incompressible fluid and an incompressible flow. The former is a thermodynamic property of the fluid expressed as $dp/d\rho=\infty$, while the latter means $dp/d(\rho v^2)\gg1$.}. Note that in conventional fluid dynamics, the notion of incompressibility is usually associated with the idea that ${\rm div}\,{\bf v}=0$, where ${\rm div}$ is the covariant divergence. But that is not true in general terms. A (necessarily non-inertial) observer, whose metric tensor is time dependent, will not, in general, report that ${\rm div}\,{\bf v}=0$, though it will report that $\dot{\cal L}_{\bf v}(\mathfrak{V})=0$ when observing an incompressible fluid. This is not a pathological condition. For instance, a fluid with a free surface can be very naturally studied with so-called $\sigma-$coordinates, whereby the free surface is described by a fixed coordinate. For such an observer, $\sqrt{g}=h(x,y,t)$, where $h$ is the function that describes the position of the surface as seen by an inertial observer.   \\


 \subsection{Kinematics}
  The trajectories of ideal Lagrangian drifters define the vector field $\mathrm{d}{x}^i/\mathrm{d}t={ v}^i$ on the space tangent to the manifold. To ${\bf v}$ we associate a flow in spacetime $\bm{\partial/\partial t}+{\bf v}$, whereas on the manifold, where we have an Euclidean structure, we can associate the action\footnote{The name stems from the fact that modulo $T$, $\lambda$ is the action per unit mass. } $1-$form $\lambda\equiv\mathbf{v}^\flat$ via musical  isomorphism. The exact (and thus closed) 2-form $\Omega\equiv d\lambda$ encodes, via Stokes theorem applied to the action integrated over closed loops (that is, the circulation), the vorticity. Unless otherwise state, when $\mathbf{v}$, $\lambda$ and/or $\Omega$ appear in the same equation, it will be understood that $\Omega=d\lambda$ and $\lambda=\mathbf{v}^\flat$.
The $(n-1)-$form $\Phi$,  the flux form, can be defined as
\begin{equation}
\Phi\equiv i_{\bf v}\mathfrak{V},
\label{eq:Sigmadef}
\end{equation}
where ${\mathfrak{V}}$ is the volume element n-form, or equivalently as 
\begin{equation}
\Phi=\star(\mathbf{v}^\flat).
\end{equation}

Since we assume \textit{ab initio} that $\mathbb{M}$ is orientable, we will use the following basis of the graded algebra 
in three dimensions
\begin{equation}
(1,dx^1,dx^2,dx^3,dx^1dx^2,dx^2dx^3,dx^3dx^1,dx^1dx^2dx^3). 
\end{equation}
(Note the cyclic nature of the index order). Also, in keeping with standard practice, we omit $\w$ when taking the exterior product of differentials (as there is no other meaning for the product). 
Here, the $x^i$'s are arbitrary coordinates. To completely characterize the coordinates, we need to specify the metric tensor $g_{ij}$, so that we the volume form is 
$\mathfrak{V}\equiv\sqrt{|g|}dx^1dx^2dx^3$.

\subsection{Dynamics}
  
One of the canonical ways to build a theory of fluids is to start from conservation of momentum. This requires vector-valued forms. 
Here, we shift the focus to vorticity, 
the advantage being that vorticity can be encoded with a scalar-valued 2-form. 
Kelvin's theorem implies that there  exists a set of observers for which kinematics alone cannot change the circulation along a close path,  a statement about vorticity. We will show how to extend this law to all observers later. For now, 
we take Kelvin's theorem axiomatically, i.e. we assume that at least for some observers it holds, so that we can write  
\begin{equation}
\dot{\cal L}_{\bf v}(\Omega)=\Gamma
\label{eq:vorticity}
\end{equation}
where the 2-form $\Gamma$ may depend only on  thermodynamic variables.  For simplicity, we limit to the simplest possible thermodynamic system close to equilibrium. 
 In this case, we know from the state postulate of thermodynamics that the state is completely specified by two independent intensive quantities. Here by independent we mean that in general changes in one variable do not necessarily imply in changes in the other. We do not mean independent in the geometric sense (i.e. the exterior product of their exterior derivatives can be zero). Being intensive, we describe them with 0-forms. 
 We choose the pressure $p$ as one of the two, and let the other be denoted with $b$.  The precise nature of $b$ will determine the appropriate equation of state.
Regardless, 
the most general expression for $\Gamma$ is\footnote{More precisely, this is the expression of the pullback $\phi^*\Gamma$, where $\phi: {\bf x} \to (p({\bf x}),b({\bf x}))$, to the thermodynamic space.} 
\begin{equation}
\Gamma=h(b,p)dbdp.
\label{eq:Gammapb}
\end{equation}
It follows immediately that $\Gamma$ is a closed 2-form, and by applying $d$ to both sides of (\ref{eq:vorticity}) we see that if $\Omega$ is closed at $t=0$, it remains so at any later time.
The converse of Poincar\`e lemma ensures that we can write\footnote{The above statement is valid locally. Globally, it depends on the topology of the manifold, as required by de Rahm's cohomology theory. The topology of the manifold may of course change with time, e.g. if singularities develop: this is related to the existence and uniqueness of solutions. We are staying well clear of that for now.} 
\begin{equation}
\Gamma=d(g(b,p)dp-df),\,g(b,p)=\int^bh(s,p)ds,
\end{equation}
where $f$ is an arbitrary 0-form. Then, from (\ref{eq:vorticity}) and the properties of the Cartan derivative, we can immediately write an equation for the action  
\begin{equation}
\dot{\cal{L}_{\bf v}}(\lambda)=-df'+g(b,p)dp.
\end{equation}
The term $df'$ includes the effects of a conservative external field with potential ${\cal Z}$, and therefore we can write it 
as $f'={\cal Z}+f''$, where $f''$ is an unknown scalar function with the property that $f''=0$ for a fluid at rest.
For  a non-interacting gas of free particles (a pressure-less inviscid gas (thus $p=0$) in free space, so that ${\cal Z}=0$)  
the rate of change  of the action form $\lambda-E_kdt$ (with $E_k\equiv i_{\lambda^\sharp}\lambda/2$) {\em in } $\mathbb{G}$ \footnote{In an inertial space-time frame, which is where Kelvin's theorem holds}  must be $\dot{d}E_k$ \citep{flanders_differential_1989}, implying 
that  $\dot{\cal L_{\bf v}}(\lambda)=dE_k$, which is satisfied if $f''=-E_k$.
At rest, we postulate a hydrostatic balance, i.e., $dp=-\rho d{\cal Z}$, and thus
\begin{equation}
\rho g(b,p)=-1, 
\end{equation}
that is $g(b,p)=-\rho^{-1}$. 
Therefore, under the following \textit{a priori} assumptions: 
\begin{enumerate}
\item Kelvin's theorem;
\item The State Postulate; 
\item Conservation of energy in free systems;
\item Hydrostatic balance at rest;
\end{enumerate}
the action  equation  assumes the following form
\begin{equation}
{\ldot_{\bf v}(\lambda)=d(E_k-{\cal Z})-\frac{dp}{\rho}.}\label{eq:lambda}
\end{equation}
The first term on the r.h.s. is the differential of the Lagrangian per unit mass.
To facilitate comparison with the vector-calculus formulation, we can rewrite (\ref{eq:lambda}) as 
\begin{equation}
\frac{\partial\lambda}{\partial t}=-i_{\bf v}\Omega-d({\cal Z}+E_k)-\rho^{-1}dp.
\label{eq:lambda1}
\end{equation}
 Expressed on a Cartesian basis, (here and thereafter $(\cdot)_{,j}\equiv \partial\cdot/\partial x^j$ )
 \begin{equation}
 \begin{split}
i_{\bf v}\Omega=i_{\bf v}((v_{2,1}-v_{1,2})dx^1dx^2+(v_{3,2}-v_{2,3})dx^2dx^3+(v_{1,3}-v_{3,1})dx^3dx^1)\\
=(v^3(v_{1,3}-v_{3,1})-v^2(v_{2,1}-v_{1,2}))dx^1+(v^1(v_{2,1}-v_{1,2})-v^3(v_{3,2}-v_{2,3}))dx^2\\+(v^2(v_{3,2}-v_{2,3})-v^1(v_{1,3}-v_{3,1}))dx^3
=v^iv_{j,i}dx^j-E_{k,j}dx^j,\,
\end{split}
\end{equation}
which substituted  in (\ref{eq:lambda1}) gives 
\begin{equation}
\left(\frac{\partial v_j}{\partial t}+v^iv_{j,i}+{\cal Z}_{,j}+\frac{p_{,j}}{\rho}\right)dx^j=0.
\end{equation}
To be identically zero on the manifold, the components (i.e., the bracketed terms) must be zero, and we recover the standard Euler equations for the Cartesian components of the velocity vector.

To complete the description, we need a constitutive relationship 
\begin{equation}
{f(p,\rho,b)=0}\label{eq:state}
\end{equation}
which depends on the choice of $b$. 
Finally, we need an equation for $b$, which we can take of the form
\begin{equation}
{\ldot_{\bf v}(b)=S,}\label{eq:thermo}
\end{equation}
where $S$ may represent internal processes. In simple cases, (e.g., an isentropic or isothermal fluid) we can set $S=0$ in which case $b$ is a LI. 

Eq. (\ref{eq:CAI},\ref{eq:lambda},\ref{eq:state},\ref{eq:thermo}) constitute a set of prognostic equations for the 4 unknowns. 
\subsubsection{Boundary conditions}
Let the $n-1$ dimensional solid boundary be described by a (set) of functions $S(i): (x^1,\ldots,x^{n-1})\to (y^1,\ldots,y^n)$. The no-flux condition becomes a condition on the pullback of the flux form $S^\star\Phi=0$.

\subsection{Some conservation laws}
 The algebraic nature of  exterior calculus makes the derivations of conserved quantities very straightforward. 
 Few passages are involved, and in a truly coordinate-free way (i.e., the results are guaranteed not to be dependent on a particular choice of coordinates).
\subsubsection{Energy $n-$forms}
Assuming that the metric tensor is time independent, 
we apply $i_{\bf v}$ to both sides of  (\ref{eq:lambda}). It is a matter of  straightforward algebraic manipulations  (recall that the Lie derivative commute with the interior product of the field, and $i_{\bf v}\lambda=2E_k$) to show that 
\begin{equation}
\dot{\cal L}_{\bf v}(E_k)=-\dot{\cal L}_{\bf v}({\cal Z})-\rho^{-1}{\cal L}_{\bf v}(p)+\frac{\partial{\cal Z}}{\partial t},
\end{equation}
and multiplying both sides by the mass 3-form CAI defined in (\ref{eq:CAI})
\begin{equation}
\dot{\cal L}_{\bf v}(\mathfrak{E}_k+\mathfrak{E}_p)=-{\cal L}_{{\bf v}}(\mathfrak{p})+pd\Phi+\frac{\partial{\cal Z}}{\partial t}\mathfrak{M}.
\label{eq:KE_PE}
\end{equation}
Here $\mathfrak{E}_{k}\equiv E_k\mathfrak{M}$  and $\mathfrak{E}_p\equiv {\cal Z} \mathfrak{M}$ are  the  Kinetic and Potential energy n-forms, and $\mathfrak{p}\equiv p\mathfrak{V}$ is the  energy  $n-$form associated to the pressure. As expected, even in the adiabatic case considered here, 
the sum of Kinetic and Potential energy is not a CC. Introducing the internal energy $\mathfrak{E}_I$,  which satisfies in the adiabatic case 
\begin{equation}
\ldot_{\bf v}(\mathfrak{E}_I)=-pd\Phi
\end{equation}
(essentially, this is $pdV$ from elementary thermodynamics), we
arrive at 
\begin{equation}
\ldot_{\bf v}(\mathfrak{E}_k+\mathfrak{E}_p+\mathfrak{E}_I)=-{\cal L}_{\bf v}(\mathfrak{p})+\frac{\partial{\cal Z}}{\partial t}\mathfrak{M} \label{eq:Bernoulli}
\end{equation} 
If the external potential is time independent, then the sum of kinetic, potential and internal energy, is a CC, which in steady flows becomes a MCC with the addition of the pressure form. In the latter case (steady flow in steady external potential), using Proposition III from the previous section, we obtain immediately that the Bernoulli function 
\begin{equation}
B\equiv E_k+{\cal Z}+E_I+\rho^{-1}p,
\end{equation} 
is a LI. 

%
\subsubsection{Potential vorticity}
Let us assume that  one of the thermodynamic variables (let it be $s$)  is a LI, i.e. $\dot{\cal L}_{\bf v}(s)=0$. Exterior multiplying (\ref{eq:vorticity}) by $ds$, and using the fact that by the state postulate $ds\w\Gamma=0$,  we  have
\begin{equation}
\begin{split}
0=ds\w\dot{\cal L}_{\bf v}(\Omega)=\dot{\cal L}_{\bf v}(ds\w\Omega)-\dot{\cal L}_{\bf v}(ds)\w\Omega=
\dot{\cal L}_{\bf v}(ds\w\Omega)-d(\dot{\cal L}_{\bf v}(s))\w\Omega=\dot{\cal L}_{\bf v}(ds\w\Omega).
\end{split}
\label{eq:PVder}
\end{equation}
Here we have fastidiously  written all the passages to indicate how effortless the derivation is. On a three-dimensional manifold, the 3-form $\mathfrak{P}\equiv ds\w\Omega$ is a MCC, 
and therefore the scalar $q\equiv \star\mathfrak{P}/\star\mathfrak{M}$, which is called with a particularly unfortunate choice of words Potential Vorticity (PV), is a LI. Not to be overlooked is that fact that (\ref{eq:PVder}) holds as long as the dimension of the manifold is equal to or greater than three, but it can be associated to a LI only in a three-dimensional space.
\subsection{Helicity}
We know that $\Gamma$ in (\ref{eq:vorticity}) must be closed, and thus locally exact by the converse of Poincar\'e lemma. Let $\Gamma=d\gamma$.
If we right multiply (\ref{eq:lambda}) by $\Omega=d\lambda$ and left multiply 
(\ref{eq:vorticity}) by $\lambda$, we obtain 
\begin{equation}
\dot{\cal L}_{\bf v}(\lambda\w\Omega)=d(\lambda\w\gamma).
\end{equation}
The 3-form $\lambda\w\Omega$ is called the helicity. If the flow is barotropic, i.e. $\gamma=dg$ for some $0-$form $g$, then the r.h.s of the above equation is not only exact (this must be expected, as every $n-$form  via Hodge decomposition is exact\footnote{Assuming the manifold is not overly pathological.}), but can be written as $-d(g\Omega)$.  
Hence, if  $V$ is a volume centered on a region where vorticity is localized, i.e. $\Omega=0$ on the boundary $\partial V$, 
the helicity integrated over $V$  is conserved. Because of the r.h.s., the Helicity is not a CC, hence it is not possible in 3-D to associate a scalar LI to the helicity as we did for the potential vorticity. 
\subsection{A final comment}
In the last few sections we have derived several conservation laws. The readers should now appreciate how natural the derivation of these laws is. It is usually a matter of hitting the equations
with $d$'s or $i_{\bf v}$'s and carry out almost embarrassingly trivial exterior algebraic manipulations. For the sake of comparison, 
contrast how PV conservation is derived with exterior calculus (Eq.~\ref{eq:PVder}) with the derivation based on vector calculus \citep[see, e.g., ref.][p. 38--43]{pedlosky_geophysical_1986}. 
Even spelling out {\em all} passages, it takes {\em one} line to show that $ds\w\Omega$ is an MCC with the present formalism, as opposed to one full page using vector calculus, not to mention the fact that in Pedlosky a crucial (and cumbersome) part of the derivation is left to the reader.  It is telling that to give a physical interpretation of the derivation, Pedlosky with considerable acumen, resorts to a geometrical interpretation based implicitly on a volume integral, \textit{de facto} recognizing the 3-form nature of PV. With exterior calculus, this interpretation is self evident  (it is a 3-form!). 
Of course, an argumentative reader may retort that  to derive conservation of PV in one line requires a (modest) mastery of  exterior calculus, which shifts the intellectual cost elsewhere. 
However, while the intellectual resources expended in deriving PV conservation cannot be "recycled" elsewhere, exterior calculus is a powerful general purpose tool, which requires a one-time investment but can be used time and again in a variety of settings.  Moreover, 
while it is true that exterior calculus is by all account a very profound way to look at geometry,
the rules of symbolic manipulation (that is, its algebra \textit{sensu lato}) are rather 
easy to master and very intuitive. 

\section{Observers in inertial and non-inertial frames}

In  (\ref{eq:lambda}) the apparatus of exterior calculus is restricted to the spatial dimensions, 
and as such 
this law involves elements of the graded algebra over a manifold $\mathbb{M}^n$ and its tangent space. 
Eq.~(\ref{eq:lambda}), and any other equation written with exterior calculus objects, (e.g., (\ref{eq:CAI})), is coordinate-free in the following sense: if $S$ is a change of coordinates in the normal sense 
\begin{equation}
S: \hat{\mathbb{M}}\to \mathbb{M},  x^i= x^i(\hat x^1,\ldots,\hat x^n), \label{eq:S}
\end{equation} then the pullback $S^*$ acts in a transparent way on the operators, i.e.,
\begin{equation}
S^*\ldot_{{\bf v}}(\lambda)=\ldot_{\hat{\bf v}}(S^*\lambda), \,S^*d(\alpha^p)=d(S^*\alpha^p),\ldots
\end{equation}
 Therefore exterior calculus provides a description which is abstracted from any particular coordinate system that we may use to describe the spatial manifold. This is what we mean by a coordinate-free description.

However, the stage of classical physics is $\mathbb{R}\times\mathbb{M}$. Consider an observer whose specific set of charts (the manifold may need more than one chart to cover it, though in the following for simplicity we will assume that one chart suffices) that cover $\mathbb{M}$ do not depend on time, together with a metric tensor (we need the latter to "measure" things).
Then from a single "observer", we can generate a set of observers applying transformations like (\ref{eq:S}). Note that such transformations are a subset of EGTs, which do not involve time. We call this {\em improper} EGTs. 
We call a set of observers generated in such a way  a \textit{frame}.  
In other words, {\em a frame is a collection of observers such that if ${\cal A}$ and ${\cal B}$ are observers belonging to the frame, then 
$\hat{x}^i=\hat{x}^i(x^1,\ldots,x^n)$, where the $\hat{x}^i$'s are the coordinates of observer ${\cal A}$ expressed in terms of the coordinates of observer ${\cal B}$.} 
The principle of covariance \citep[see ref. ][p. 11 for a formal definition]{cantwell_introduction_2002} requires that all observers within a frame, by means of experiments (i.e. quantitative measurements which are not possible without a coordinate system), must arrive at the same coordinate-free formulation of physical laws. The principle of covariance does not however require that observers in different frames arrive at the same physical laws.
The principle of Galileian relativity applied to fluid dynamics requires {\em the existence of a set of privileged frames, the inertial frames}. Observers belonging to an inertial frame  conducting experiments on fluids will arrive at (\ref{eq:lambda}) and {\em a fortiori} to (\ref{eq:vorticity}).   
\subsection{Observers in different frames}
By definition, observers that belong to different frames (inertial or not) are linked by coordinate transformations that depend non trivially on time,  which we call {\em proper} EGTs. 
Whereas  restricting the machinery of exterior calculus to the spatial coordinates alone 
ensures a coordinate-free description for observers that belong to the same frame, the situation is more complicated when considering different frames. In this case, we need to apply exterior calculus over $\mathbb{G}$ applied to forms that belong to $G/T$.  For simplicity, we assume in the following that the spatial manifold can be mapped by a single chart. Manifolds requiring multiple charts do not introduce qualitative differences. 

Assume that observer ${\cal A}$ belonging to an {\em inertial} frame $A$ is a set of coordinates $x^\alpha$'s, and likewise observer ${\cal B}$ in an {\em arbitrary} frame B is a set of $\hat x^\alpha$'s. We must be able to write   
\begin{equation}
\hat x^\alpha=\hat x^\alpha(x^0,\ldots,x^n), \alpha=0,\ldots,n. 
\end{equation}
Being  classical space-time observers, they must be linked by a (proper) EGT, i.e. $\hat x^0=x^0+\mathrm{const.}$. Recall that the Cartan derivative is the Lie derivative 
in $\mathrm{G}/T$, and it measures the rate of change of forms under the action of a flow. For a given coordinate system, the flow in $\mathbb{G}$ is described by its group operators $u^\alpha\bm {\partial/\partial x^\alpha}$, where $u^0=1$. Simple application of the chain rule gives
\begin{equation}
\hat u^\beta\bm{\frac{\partial}{\partial\hat x^\beta}}=u^\alpha\frac{\partial\hat x^\beta}{\partial x^\alpha}\bm{\frac{\partial}{\partial\hat x^\beta}},
\end{equation}
that is the infinitesimals change as contravariant vectors (no surprises here, as they belong to the tangent space). Note that since the observers are related by an EGT, $\hat u^0=1$ so $\hat{\bf u}$ is a flow. 
At this point we define the null flow for observer ${\cal A}$ as the flow whose infinitesimals $(1,u^{*1},\ldots, u^{*n})$ satisfy
\begin{equation}
u^{*j}\frac{\partial \hat x^i}{\partial x^j}=-\frac{\partial\hat x^i}{\partial x^0}. \label{eq:Comp_vel}
\end{equation}
As can be easily verified, the null flow is the vector field tangent to the trajectories of drifters in ${\cal A}$ that is mapped by the EGT to the flow whose infinitesimals are $(1,0,\ldots,0)$, so that the same drifters in ${\cal B}$ appear stationary\footnote{Alas, in classical space time we cannot stop time!}. 
The interesting twist occurs now. Suppose that ${\cal A}$ measures  a flow with infinitesimals 
\begin{equation}
u^i=u'^i+u^{*i}.
\end{equation}
The infinitesimals for ${\cal B}$ are 
\begin{equation}
\hat u^i=u'^j\frac{\partial \hat x^i}{\partial x^j},\label{eq:infhat}
\end{equation}
which is not surprising, as (\ref{eq:Comp_vel}) is nothing but classical  law of composition of velocities. 
The action in ${\cal A}$ is  
\begin{equation}
\lambda=(u'_i+u^*_i)dx^i.
\end{equation}
Modulo $T$, the action under the pullback $S$ given by the EGT connecting the two coordinate systems becomes 
\begin{equation}
S^*\lambda=\left(u_j\frac{\partial x^j}{\partial\hat x^i}+u^*_j\frac{\partial x^j}{\partial \hat x^i}\right)d\hat x^i=\hat\lambda+\lambda^f.
\end{equation}
The twist is that while the infinitesimals of the null flow are mapped to zero, the svtion $\lambda^f$ associated to the null flow is not zero under the action of the pullback modulo $T$! 
Wrapping up, under the pullback to the space-time coordinates of ${\cal B}$, (\ref{eq:lambda}) becomes (hats applied to $0-$forms denote the $0-$forms in the new coordinates)
\begin{equation}
\ldot_{\bf{\hat v}}(\hat \lambda)=-\ldot_{\bf{\hat v}}(\lambda^f)-\frac{d\hat p}{\hat\rho}-d(\hat {\cal Z}-E_k),
\end{equation}
where $E_k=\frac{1}{2}(\hat E_k+ i_{(\lambda^{f\sharp})}\lambda^f+u'_iu^{*i}+u'^iu^*_i)$,  and the infinitesimals of the group operator ${\bf\hat v}$ are given by (\ref{eq:infhat}). After simple manipulations 
we arrive at the final form of (\ref{eq:lambda}) under the pullback (in space-time!)
\begin{equation}
\ldot_{\bf{\hat v}}(\hat \lambda)=\left[-\frac{\partial}{\partial t}\lambda^f-i_{\bf \hat v}d\lambda^f\right]-\frac{d\hat p}{\hat\rho}+d\left(\hat E_k-\frac{1}{2}i_{(\lambda^{f\sharp})}\lambda^f-\hat{\cal Z}\right). \label{eq:lambda_EGT}
\end{equation}

Comparing (\ref{eq:lambda1}) with (\ref{eq:lambda_EGT}) shows that, in an arbitrary frame, the change in action along a path is due, in addition to the effect of material causes encoded in the potential ${\cal Z}$ and changes in thermodynamic state, to the effect of the bracketed terms in (\ref{eq:lambda_EGT}) containing $\lambda^f$, which we call, for reasons that will soon become clear, the generalized Coriolis force. Also, the kinetic energy of the frame appears as a ``potential'' energy.

We call $\lambda^f$ the frame action, since it is a property of the frame, not of the coordinates. 
This suggests a classification of frames and the proper EGT's connecting them based on the form of the equations. In the following, A is an inertial frame. 
\begin{enumerate}
\item Inertial or Galileian frames: These are frames where (\ref{eq:lambda}) holds. In other words, $\lambda^f$ is exact, constant in time, and $i_{(\lambda^{f\sharp})}\lambda^f$ is closed, i.e. constant. The most general proper EGT that leaves (\ref{eq:lambda}) invariant 
is a Galileian boost (an $n$-dimensional subgroup of the  $n!/(2(n-2))!+2n+1$-parameter group of Galileian transformations\footnote{The group is made up of $n+1$ coordinate translations in space-time, $n$ Galileian boosts and $n!/(2(n-2))$ coordinate rotations.}) Thus, if we \textit{know} that a frame A is inertial (based on the definition above), we can generate \textit{all} of the inertial (or Galileian) frames by application of  Galileian boosts to an observer in A.

\item Pseudo-inertial or Maxwellian frames:  These are frames 
 such that $\lambda^f=dw$ is an exact 1-form.  Under the proper EGT that connects ${\cal A}$ to a Maxwellian frame, the generalized Coriolis force is exact, and thus can be absorbed 
 into a pseudo-potential.  The vorticity equation is invariant under pseudo-inertial EGTs. 

\item Leibnizian frames: Frames that can be accessed from an inertial frame via EGTs such that $\lambda^f$ is not closed. In this case, we call $d\lambda^f$ the frame vorticity. Example of these EGTs are rigidly rotating frames (the rigidity implies that the metric does not change in time).  Both  (\ref{eq:lambda}) and (\ref{eq:vorticity})  are not invariant under proper EGTs accessing Leibnizian frames.

\end{enumerate}

Maxwellian frames form a subset of Leibnizian frames, and  Galileian frames are a subset of Maxwellian frames. 
From a fluid dynamic point of view, Leibnizian frames, and by extension all frames such that $d\lambda^f\neq0$ can be unequivocally distinguished. Indeed, \citet[][p. 271]{stommel_introduction_1989} describe a laboratory apparatus  (called a Compton generator)
that can be used to measure $d\lambda^f$. An observer in an enclosed box measuring a non zero $d\lambda^f$ will identify his frame as Leibnizian. 
The distinction between Maxwellian and Galileian frames is more subtle. It hinges on being able to discriminate on what causes the potential ${\cal Z}$. Consider making experiments within a closed box. 
In order to discern  a material cause (a gravitational field generated by a distribution of masses) for ${\cal Z}$ from an effect of the frame  (assuming equivalence of inertial and gravitational mass) the observer in the Maxwellian frame needs to know the complete distribution of masses in the Universe. But, with this knowledge, the observer can calculate the center of mass of the Universe,  (Mach's definition of an "absolute" frame) and with it determine her motion relative to it.   

\subsubsection{Examples}

In this section, we will consider two examples of Leibnizian.  The coordinates of an observer within a Galileian frame will be "hatted", while "hat-free" variables will denote the coordinates of an observer in the non-inertial frame.

\paragraph{Rotating layered  coordinates}
Many geophysical flows  are not far from a state of solid body rotation. In this case, it is expedient to study them from the point of view of a Leibnizian frame rotating with the fluid. Consider the  proper EGT which connects the cylindrical coordinates $(\hat x^1,\hat x^2,\hat x^3)$ of an inertial frame to the coordinate $(x^1,x^2,x^3)$ of a Leibnizian frame
\begin{gather}
\hat{x}^1=x^1\,\mbox{(radial}\,\mbox{coordinate)},\\
\hat{x}^2=x^2+\omega t\,\mbox{(azimuthal}\,\mbox{coordinate)} ,\\
\hat{x}^3=\zeta(x^1,x^2,x^3,t)\,\mbox{(vertical}\,\mbox{coordinate)}.\label{eq:vertical}
\end{gather}
Relative to the Galileian frame, this Leibnizian frame rotates counter-clockwise (assuming $\omega>0$) around the $\hat x^3$ axis and measures the third coordinate relative to a surface, which, in the Galileian frame, may not be stationary. Several types of coordinates can be represented this way, e.g. isopycnic coordinates if the surface is a surface of constant density or so-called $\sigma-$coordinates if the surface corresponds to a topographic boundary.
To calculate the frame action, we begin with the null velocity in the inertial frame, that is the solution to 
\begin{equation}
\hat{u}^{*j}\frac{\partial x^i}{\partial \hat x^j}=-\frac {\partial x^i}{\partial t}. 
\end{equation}
Eq.~(\ref{eq:vertical}) gives $x^3$ implicitly via $\hat x^3-\zeta(\hat{x}^1,\hat{x}^2-\omega t,x^3,t)=0$, so that 
\begin{equation}
\frac{\partial x^3}{\partial\hat x^1}=-\frac{\zeta_{,1}}{\zeta_{,3}},\,\frac{\partial x^3}{\partial \hat x^2}=-\frac{\zeta_{,2}}{\zeta_{,3}},\,\frac{\partial x^3}{\partial \hat x^3}=\frac{1}{\zeta_{,3}},\,\frac{\partial x^3}{\partial t}=\omega\frac{\zeta_{,2}}{\zeta_{,3}}-\frac{\zeta_{,t}}{\zeta_{,3}},
\end{equation}
and  the null velocity is 
\begin{equation}
\hat u^{*1}=0,\,
\hat u^{*2}=\omega,\,
\hat u^{*3}=\zeta_{,t},
\end{equation}
where $\zeta_{,j}\equiv\partial\zeta/\partial \hat x^j$ and $\zeta_{,t}\equiv\partial\zeta/\partial t$.
The frame action in the Leibnizian frame is  given by (we use the fact that the metric in the Galileian frame is diagonal with diagonal $(1,(\hat x^1)^2,1)$)
\begin{equation}
\lambda^f=\hat{u}^{*l}\hat{g}_{li}\frac{\partial\hat{x}^i}{\partial x^j}dx^j=
\omega (x^1)^2dx^2+\zeta_{,t}d\zeta.
\end{equation}  
 The frame vorticity is then
\begin{equation}
d\lambda^f=2\omega x^1dx^1dx^2+d\zeta_{,t}\w d\zeta\label{eq:frame_vort_iso}
\end{equation}
while the time derivative of the frame action
\begin{equation}
\frac{\partial\lambda^f}{\partial t}=d\left(\frac{(\zeta_{,t})^2}{2}\right)
+\zeta_{,tt}d\zeta
\end{equation}
Up to this point, we have not made any assumption on $\zeta(x^1,x^2,x^3,t)$ or the nature of $x^3$. In particular, if $\zeta_{,t}=0$, i.e. the reference surface is stationary, 
then using (\ref{eq:frame_vort_iso}) the generalized Coriolis force
\begin{equation}
-i_{\mathbf{v}}d\lambda^f=2\omega x^1(v^2dx^1-v^1dx^2),\label{eq:frame_vort_Cor}
\end{equation}
coincides with  the standard expression of the Coriolis force.  
To write the equation for mass conservation, we begin by observing that the volume $3-$form in $(x^1,x^2,x^3)$ coordinates is
\begin{equation}
\mathfrak{V}=|\zeta_{,3}|x^1dx^1dx^2dx^3. 
\end{equation}
A useful choice of coordinates in large-scale geophysical flows is obtained when the density is used as an independent coordinate. 
With this choice, the 
 mass $3-$form becomes 
\begin{equation}
\mathfrak{M}=x^3\mathfrak{V}.
\end{equation}
 Let ${\bf v}=v^i\bm{\partial/\partial x^i}$. Mass conservation then requires
\begin{equation}
\begin{split}
\ldot_{\mathbf{v}}(\mathfrak{M})=x^3\ldot_{\bf v}(\mathfrak{V})+\mathfrak{V}\ldot_{\bf v}(x^3)=\\
[Q_{,t}+(x^1)^{-1}(x^1Qv^i)_{,i}]x^1dx^1dx^2dx^3=0.
\end{split}
\end{equation}
where $Q\equiv -\zeta_{,3}x^3$,  
which gives a prognostic equation for $Q$.\footnote{In order for the coordinate transformation to be well defined, $\zeta_{,3}\neq 0$, and thus it must be either positive or negative. In geophysical applications, density decreases with height, so that $|\zeta_{,3}|=-\zeta_{,3}.$} 

The heaving of the isopycnals, or, more generally, of the reference surfaces, introduces an extra term in Coriolis force.  
When the above equations are restricted to geophysical applications characterized by  horizontal scales  much larger than the vertical scale (thin layer approximation), and when the motion occurs on time-scales much longer than $(\omega^{-1})$ (subinertial motion), it is possible to ignore most terms. In the action equation, the dominant balance for the component along $dx^3$ is hydrostatic, that is 
\begin{equation}
p_{,3}=-gx^3\zeta_{,3}=gQ,
\end{equation}
where $g$ is the gravitational acceleration and $p$ the pressure. Under the same approximations, $\zeta_{,tt}/g\ll 1$, therefore the time derivative of the frame can be ignored. Finally, as long as $v^3\zeta_{,3t}/g\ll 1$, and $\zeta_{,1}\zeta_{,2t}/\omega\simeq \zeta_{,2}\zeta_{,1t}/\omega\ll1$, the contribution of the heaving isopycnals to the frame vorticity can be neglected.
It is only when going beyond the thin layer approximation (e.g., going beyond the hydrostatic approximation, which necessarily implies moving toward supra-inertial frequencies), that the heaving isopycnals contribution of the frame vorticity should be included.  

\paragraph{The fluid frame of a barotropic fluid}
As a final example of Leibnizian frames we consider the frame of the fluid itself, that is, the Lagrangian frame. In this frame, obviously, the fluid appears motionless, though not necessarily steady. Consider a barotropic fluid, so that $dp/\rho=dg$ for some $0-$form g. In the fluid frame 
\begin{equation}
\frac{\partial\lambda^f}{\partial t}=-d(g+{\cal Z}+E^f_k),
\end{equation}
where $E^f_k$ is the kinetic energy of the frame. For an observer on the fluid frame, the rate of change of the frame circulation is exact. That does not mean that the frame circulation is closed. It only means that in the fluid frame, the frame vorticity is constant in time. Hodge decomposing the frame circulation 
\begin{equation}
\lambda^f=dh+d^\star\Phi,
\end{equation}
we have that both $\Phi$ (the vector potential, in vector parlance) is steady, whereas the scalar potential $h$ satisfies
\begin{equation}
h_{,t}+g+{\cal Z}+E^f_k=0.
\end{equation}
 This is the unsteady version of Bernoulli theorem. Remember that here $h$ is the scalar potential as calculated  within the fluid frame. Whereas in a generic Maxwellian frame the unsteady form of Bernoulli theorem holds only for barotropic  irrotational flows, in the fluid frame itself the Bernoulli theorem holds for barotropic flows bar none, though the frame kinetic energy includes contributions from both the rotational and solenoidal components. 
\section{Conclusions}

In this paper we formulated  a theory of Euler fluids using the language of exterior calculus. The complete set of equations in an arbitrary 
frame is
\begin{gather}
\ldot_{\bf v}(\mathfrak{M})=0\,\,\mbox{(Mass conservation)},\\
\ldot_{\bf v}(\lambda)=d\left(E_k-\frac{1}{2}i_{(\lambda^{f\sharp})}\lambda^f-{\cal Z}\right)-\rho^{-1}dp -\frac{\partial\lambda^f}{\partial t}-i_{\bf v}d\lambda^f\,\,\mbox{(Conservation of action)},\\
\ldot_{\bf v}(b)=S \,\,\mbox{(Thermodynamics)},\\
f(p,\rho,b)=0, \,\,\mbox{(State equation)},\\
\lambda={\mathbf v}^\flat,\,\,\rho=\star\mathfrak{M} \,\,\mbox{(Constitutive relationships)}.
\end{gather}
Here, $\ldot_{\bf v}(\cdot)$ is the Cartan derivative, which specializes the Lie derivative to the subset of forms in classical space-time that encode physical information, and $\lambda^f$ is the frame action, a property of the frame. 
Relative to the same equations formulated with vector calculus this formulation provides several advantages:
\begin{enumerate}
\item The formulation is coordinate-free. The equations are transparent to change of coordinates (pullbacks). Any result, such as conservation laws, obtained manipulating  the equations using exterior calculus operators is valid in any coordinate system. The maze of indexes that is typical of the tensor approach is absent or severely curtailed, allowing the physics to stand out. 
\item Rather than scalar and vector fields, physically relevant quantities are described by elements of the graded algebra of $p-$forms.
This provide a richer space where physically distinct 
quantities (e.g. intensive vs. extensive properties) are encoded by different elements of the algebra. 
\item In the standard approach, the operator that describes the evolution of fields is the Lagrangian derivative. Except when applied to intensive quantities (i.e., $0-$forms), the Lagrangian derivative does not account for all kinematic effects, resulting in the appearance of extra terms in the equations.
In the exterior calculus approach,  the Cartan derivative, whose algebraic definition is one and the same regardless of what element of the graded algebra applies to,  accounts for all  kinematic effects. This clearly separates the physical causes that can change a quantity from  purely kinematic effects.
\item The coordinate-free nature of the equation, coupled to the algebraic nature of the approach, allows to derive results cleanly and with relatively little passages, such as  the derivation of the generalized Coriolis force that describe non-inertial effects associated to proper EGTs, a novel result. 
\item The present formulation can take advantage of developments in discrete geometry, which could lead to novel schemes to numerically solve the equations that respect the underlying symmetries.  
\end{enumerate}
Of course,  we are neither envisioning, nor advocating, a total take-over of exterior calculus. However, we think that from a theoretical point of view, ignoring the tremendous progress made by mathematicians and physicists in developing field-theoretical tools during the last century is only detrimental to fluid dynamics.  

\begin{acknowledgments}
I would like to thank Dr. Santilli for introducing me to exterior calculus, and for many fruitful discussions on forms and fluid dynamics. While this work was done mostly during my free time, I would nonetheless like to acknowledge the support I received over the years by ONR and NSF. 
\end{acknowledgments}
\appendix
\begin{figure}
\includegraphics[scale=.5]{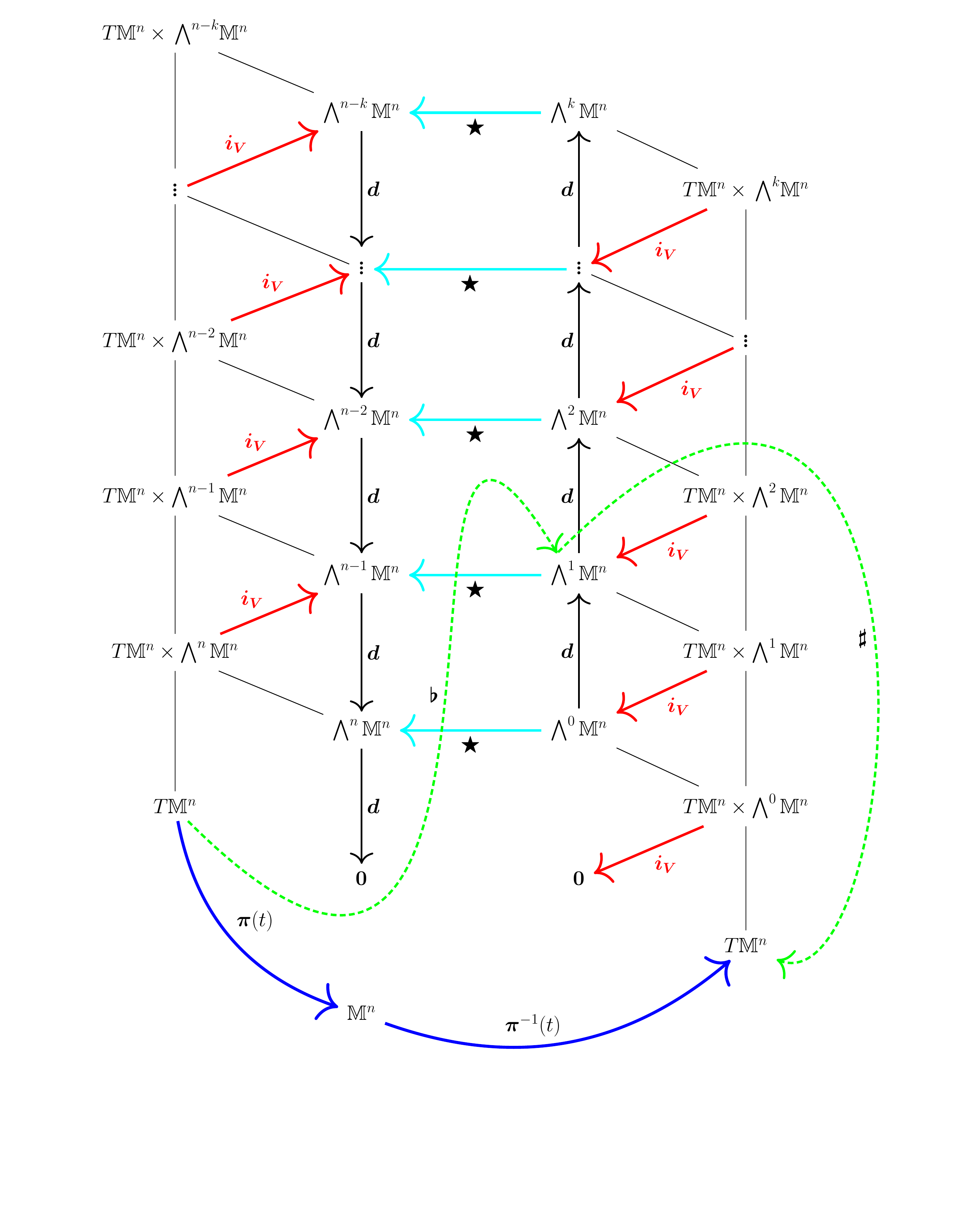}
\caption{\label{fig:App1} The graded algebra of $p-$forms over a manifold $\mathbb{M}^n$ with the operators connecting them.  Red and green arrows are isomorphisms that require a metric structure  on the manifold.}
\end{figure}
\section{"One $d$ to rule them all" (with S.M. Da Silva-Mendes). Basic definitions and properties of the operators of exterior calculus.}
In this Appendix, we give a brief overview of the objects and operators of exterior calculus. It is largely based on \citet{flanders_differential_1989}, to which the reader is referred for details.

We first define the "types" over which the operators act. The operators are then defined based on a series of properties. Readers familiar with the concept of "object oriented" programming will see a similar pattern unfolding here. Starting from a simple type (scalar functions), we "construct" new types by the repeated application of operators defined by a set of algebraic properties. Likewise, the operators are defined in term of a simple operator which is extended via a set of algebraic rules. 
\subsection{The Grassmann algebra of $p-$forms}\label{App:pforms}
Let  $\mathbb{M}^n$, denote a differentiable manifold (essentially, an object that is locally equivalent to a piece of $\mathbb{R}^n$), where $n$ is the dimension of the manifold. 
The "types" of exterior calculus are differential $p-$forms, or just $p-$forms for short. $p-$forms belong to a vector space over the real denoted with $\bigwedge^p\mathbb{M}^n$, and we call $p$ the grade of the form. As in the main article, $\alpha^p, \beta^p,\ldots$  will denote  generic $p-$forms. The direct sum 
\begin{equation}
G\equiv\overset{n}{\underset{p=0}{\oplus}}\bigwedge^p \mathbb{M}^n
\end{equation}
(note how $p$ ranges from 0 to $n$)
forms a graded algebra. Different physical quantities are encoded with with forms of an appropriate grade. 

The graded algebra is built using a bootstrapping technique, starting from $0-$forms: $\bigwedge^0\mathbb{M}^n$ is the space of smooth functions  $f:\mathbb{M}^n\to\mathbb{R}$. Algebraically, the set of smooth functions is a ring (we know how to add and multiply functions, we have the $0$ function, and the $1$ function, but in general functions need not to have an inverse relative to multiplication). 
Physically, we can think of functions as objects that assign values (temperature, salinity, entropy,\ldots) to $0-$dimensional surfaces (i.e. points) of $\mathbb{M}^n$. Consider the space that is obtained considering all possible sums of terms like $a(x^1,\ldots,x^n)df(x^1,\ldots,x^n)$, where $a$ and $f$ are smooth functions from $\mathbb{M}^n$ into $\mathbb{R}$. (i.e., $0-$forms) and $d$ in this context stands for the standard differential from elementary calculus of multivariate functions. These objects  assign a value to curves (1-dimensional surfaces) in $\mathbb{M}^n$ and make up the space of $1-$forms. It is a vector space over the ring of $0-$forms. 
All the other spaces in the graded algebra are generated via a bootstrapping process that uses $0-$forms (a ring) and $1-$forms (a vector space over the ring), together with a composition operator $\w$ (defined next) to build higher grade forms. For our particular case,  
the goal is to generate objects that assign values to $p-$dimensional surfaces in $\mathbb{M}^n$. The resulting collection is an example of a Grassmann algebra.

\subsection{Operators on $p-$forms}
Over the graded algebra  we define three types of operations. One is a binary operator (the exterior product), which is used to generate the algebra given $0-$ and $1-$forms, 
and two are unary operators, the exterior derivative and the interior product. These operations require only that the manifold $\mathbb{M}^n$ be differentiable. 
\subsubsection{Wedge or exterior product}\label{App:wedge}
This is the operation that gives the name to the whole calculus. It is also the operator that is used to construct the graded algebra once $0-$ and $1-$forms are defined. It is a binary operation $\w:\bigwedge^q\mathbb{M}^n\times\bigwedge^p\mathbb{M}^n\to \bigwedge^{(p+q)}\mathbb{M}^n$ that is associative, distributive relative to the sum of $p-$forms and in addition satisfies 
\begin{gather}
\alpha^p\w\beta^q=(-1)^{pq}\beta^q\w\alpha^p. \label{eq:A1}
\end{gather}
When applied to $1-$form, we have 
\begin{equation}
\alpha^1\w\beta^1=-\beta^1\w\alpha^1,
\end{equation}
while the associativity combined with (\ref{eq:A1}) implies
\begin{equation}
\alpha^p\w (\beta^0\w\gamma^q)=\beta^0\w(\alpha^p\w\gamma^q),
\end{equation}
where $\beta^0$ is a $0-$form, that is a real valued function on the manifold. The wedge product of two $0-$forms is the standard product of functions. 
The bootstrapping process start with  $0-$ (a ring) and $1-$forms (a vector space of the ring). Using the wedge product we construct $\bigwedge^2\mathbb{M}^n$ by requiring that it contains all wedge products of pairs $1-$forms, and all   
the linear combinations of products of $1-$forms where the coefficients are $0-$forms so that in the end we have a vector space over the ring of $0-$forms.
Next we consider all the linear combinations of exterior products of $1-$ and $2-$forms to generate $\bigwedge^3\mathbb{M}^3$ and so on. 
Since, given $n$ objects, at most we can pick $n$ distinct objects, $\bigwedge^p\mathbb{M}^n=\emptyset$ when $p>n$.
\subsubsection{Exterior derivative}\label{App:d}
In the preceding section, we showed that a graded algebra can be built using $0-$ and $1-$forms as building blocks. At this point, we know what $0-$forms are, and we used the notion of differential $df$ of a function $f$ to generate a vector space over the $0-$forms as the space that includes the differential of $0-$forms and their linear combinations, i.e. objects that can be written as 
\begin{equation}
\alpha^1=\sum_i\alpha^0_id\beta^0_i,
\end{equation}
where the $\alpha^0_i$'s and $\beta^0_i$'s are arbitrary $0-$forms.  
Having constructed the space of $0-$ and $1-$forms, we can operationally construct all other elements of the graded algebra via the bootstrapping procedure using the wedge product.

Here, we extend the idea of the differential operator from $0-$ to all other forms. To wit, the exterior derivative is the unary operator $d:\bigwedge^p\mathbb{M}^n\to\bigwedge^{(p+1)}\mathbb{M}^n$ which satisfies the following properties:
\begin{gather}
\alpha^0(B)-\alpha^0(A)=\int_A^Bd\alpha^0,\label{eq:A5}\\
d(\alpha^p+\beta^q)=d\alpha^p+d\beta^q,\,(\mbox{linearity})\label{eq:A6} \\
d(d(\alpha^p))=0,\,(\mbox{nilpotency})\label{eq:A7}\\
d(\alpha^p\w\beta^q)=(d\alpha^p)\w\beta^q+(-1)^p\alpha^p\w (d\beta^q)\, (\mbox{antiderivative}).\label{eq:A8}
\end{gather}
In (\ref{eq:A5}), the integral is on a oriented curve that joins $A$ to $B$, and it is the property that basically says that on $0-$forms $d$ is the differential. 
Indeed, from the fundamental theorem of calculus, we then know that if $(x^1,\ldots,x^n)$ are local coordinates on $\mathbb{M}^n$, then (using Einstein convention on repeated indexes)
\begin{equation}
d\alpha^0=\frac{\partial\alpha^0}{\partial x^i}dx^i,\label{eq:A9}.
\end{equation}
While (\ref{eq:A9}) can be used instead of (\ref{eq:A5}), we prefer (\ref{eq:A5}) because it is coordinate independent. Once defined on $0-$forms, the axiomatic set of properties (\ref{eq:A6}-\ref{eq:A8}) is  used to {\em uniquely} extend the operator $d$ over the entire graded algebra.  

 Since $p-$forms belong to finite dimensional vector spaces, it is only necessary to generate basis. For example, on a 3-dimensional manifold with local coordinates $x^1,x^2,x^3$ a choice of basis is given by
\begin{gather}
\bigwedge^0\mathbb{M}^3 \to 1,\\
\bigwedge^1\mathbb{M}^3 \to dx^1,dx^2,dx^3,\\
\bigwedge^2\mathbb{M}^3 \to dx^1\w dx^2,dx^2\w dx^3, dx^3\w dx^1\\
\bigwedge^3\mathbb{M}^3 \to dx^1\w dx^2\w dx^3.
\end{gather}
From now on, when considering the exterior product of simple differentials, we will omit the $\w$, as there is no other way to interpret the juxtaposition. In terms of the basis, 
$2-$ forms are $\alpha^2=\alpha^0_{ij}dx^idx^j$. Note that because of the antisymmetry of the exterior product of two $1-$forms, only the antisymmetric part of $\alpha^0_{ij}$ contributes to the sum. 

 The fundamental theorem of calculus becomes a special case of  Stokes' theorem: \textit{Let $C$ be a bounded $(p+1)-$dimensional surface with boundary $\partial C$ (a $p-$dimensional surface). Then }
\begin{equation}
\int_{\partial C} \alpha^p=\int_Cd\alpha^p.\label{Th:Stokes}
\end{equation}
Hence, $p-$forms are object that assign values to $p-$dimensional surfaces of the manifold.
We leave to the reader to verify that when $\mathbb{M}^3=\mathbb{R}^3$, $d$ acting on $1-$forms is equivalent to   the Curl, and $d$ acting on $2-$forms is equivalent to the Divergence, and thus  we see that  the Gauss-Ostrogradsky and Kelvin-Stokes theorems from ordinary vector calculus are all special cases of (\ref{Th:Stokes}). 
\subsubsection{The interior product}\label{App:iv}
Every vector space has a dual space, the space of functionals over the vector space. The dual space  of $1-$forms {\em qua} vector space is the space of vectors that, once a coordinate system $(x^1,\ldots,x^n)$ is chosen, can be written as 
\begin{equation}
{\mathbf v}=v^i\bm{\frac{\partial}{\partial x^i}}.
\end{equation}
On $1-$ forms, we define the interior product via
\begin{equation}
i_{\mathbf v}(\alpha^0_idx^i)\equiv v^i\alpha^0_i.\label{eq:A17}
\end{equation}
Just as we did for the exterior derivative, we extend axiomatically the interior product to the entire graded algebra as follows: for a given vector ${\mathbf v}$ of the dual space of $1-$forms, the interior product $i_{\mathbf v}:\bigwedge^{p}\mathbf{M}^n\to\bigwedge^{(p-1)}\mathbf{M}^n$ satisfies (\ref{eq:A17}) and is uniquely extended to the rest of the graded algebra by requiring that it be linear, nilpotent and an antiderivative. We stipulate that $i_\mathbf{v}\alpha^0=0$. 

\subsection{Euclidean manifolds}
If the manifold ${\mathbb M}^n$ possesses an Euclidean structure, i.e. a metric  $g_{ij}$ with signature $(n,0,0)$, 
then the Riesz representation theorem guarantees the existence of natural isomorphisms between $1-$forms and vectors, and between $p-$forms and $(n-p)-$forms. We introduce first the former.
\subsubsection{The musical isomorphisms}
Consider a $1-$form $\alpha^1=\alpha^0_idx^i$. Under a change of coordinates, we have
\begin{equation}
\alpha^1=\alpha^0_idx^i=\alpha^0_i\frac{\partial \tilde x^j}{\partial x^i}d\tilde x^j,
\end{equation}
which shows that the components of the $1-$form transforms like the components of a covariant vector.\footnote{ While this property is used to {\em define} what a contravariant vector is in standard tensor calculus, this is not the case here. We defined $1-$forms without any metric notion.} Thus, if we are given a covariant vector, we can associate a $1-$form to it. Similarly, the components of a vector $\mathbf{v}=v^i\bm{\partial/\partial x^i}$ transform like a contravariant vector.  Just as we use the metric to introduce an isomorphism between co- and contra-variant rank-1 tensors, we can use it here 
to introduce isomorphisms between $1-$forms and vectors on the tangent space. We thus can introduce the two so-called musical isomorphisms, 
\begin{equation}
{\bf v}^\flat=v^ig_{ij}dx^j=v_idx^i=\lambda,
\end{equation}
called the flat since it "lowers" the index, and its inverse, the sharp,
\begin{equation}
\lambda^\sharp=v_ig^{ij}\bm{\frac{\partial}{\partial x^j}}=v^j\bm{\frac{\partial}{\partial x^j}}=\mathbf{ v}.
\end{equation}

\subsubsection{The Hodge star}\label{App:HS}
We leave to the reader to prove that 
\begin{equation}
{\rm dim}\bigwedge^p \mathbb{M}^n=\frac{n!}{p!(n-p)!}={\rm dim}\bigwedge^{n-p} \mathbb{M}^n.
\end{equation}
This suggests that it may be possible to introduce a natural isomorphism between $p-$ and $(n-p)-$forms. To do that, we first introduce an inner product $(\cdot,\cdot)$ on $1-$forms as follows
\begin{equation}
(\alpha^1,\beta^1)\equiv i_{({\alpha^1})^\sharp}\beta^1=i_{({\beta^1})^\sharp}\alpha^1.
\end{equation}
Note that this definition relies on the $\sharp$ isomorphism, and thus depends on the existence of  an Euclidean structure on the manifold. 
Next, given two $p-$forms $\alpha^p=\alpha^1_1\w \ldots\w\alpha_p^1$ and $\beta^p=\beta^1_1\w\ldots\w\beta^1_p$, we write
\begin{equation}
(\alpha^p,\beta^p)=|(\alpha^1_i,\beta^1_j)|,
\end{equation}
where $|a_{ij}|$ is the determinant of the matrix whose elements are $a_{ij}$. Since the determinant is alternating multilinear, we can extend the definition to a generic pair of $p-$forms. For a given $\beta^{(n-p)}$, consider the following map
\begin{equation}
f_{\beta^{(n-p)}}:\bigwedge^p\mathbb{M}^n\to \bigwedge^n\mathbb{M}^n; f_{\beta^{(n-p)}}(\alpha^p)=\alpha^p\w\beta^{(n-p)}. 
\end{equation}
The space of $n-$forms is 1-dimensional, and we choose $\mathfrak{V}\equiv \sqrt{|g|}dx^1\ldots dx^n$ as its basis. This $n-$form is special in that integrated over a submanifold of $\mathbb{M}^n$ returns its volume.\footnote{$n-$forms can of course be integrated on manifolds that do not have an Euclidean structure.}  Therefore, $f_{\beta^{(n-p)}}$ is isomorphic to a functional on the $p-$forms, now seen as an inner product vector space, and by Riesz representation theorem there exists a unique $p-$form $\gamma^p$ such that 
\begin{equation}
f_{\beta^{(n-p)}}(\alpha^p)=(\alpha^p,\gamma^p)\mathfrak{V}.
\end{equation}
The operator that associates $\beta^{(n-p)}$ to $\gamma^p$ is called the Hodge star and we write
\begin{equation}
\star\gamma^p=\beta^{(n-p)}. 
\end{equation}
From its definition
\begin{equation}
\alpha^p\w\star\beta^p=\beta^p\w\star\alpha^p=(\alpha^p,\beta^p)\mathfrak{V},
\end{equation}
and 
\begin{equation}
\star\star\alpha^p=-(-1)^{p(n-p)}\alpha^p,
\end{equation}
which shows that modulo a factor (-1), the Hodge star is the inverse of itself. We now state a few useful results  whose verification is left to the reader.\\
\begin{itemize}
\item $\star(\alpha^1\wedge(\star\alpha^p))=(-1)^{\gamma(n,p)}i_{\bf (\alpha^1)^\sharp}\alpha^p$ with  $\gamma(n,p)=n(p+1)$. 
\item The left exterior multiplication by $\alpha^1$ of a $p-$form is equivalent 
to applying to the left $-(-1)^{n(n-p)}\star i_{(\alpha^1)^\sharp}\star$. 
\item To a vector ${\bf v}$ we can naturally associate the $n-1$ form $i_{\bf v}\mathfrak{V}=\star(\mathbf{v}^\flat)$.
\item  $i_{\bf v}i_{\bf u}(\cdot)=-i_{\bf u}i_{\bf v}(\cdot)$ and $i_{{\bf v}+{\bf u}}(\cdot)=i_{\bf v}(\cdot)+i_{\bf u}(\cdot)$.
\item  Since ${\bf v}=v^i\bm{\partial/\partial x^i}$,  we can think of 
the interior product as the operator that substitutes a $dx^i$ in a form with the corresponding $v^i$ multiplied by the appropriate $\pm1$ factor, depending on the position of $dx^i$. 
\end{itemize}
\subsubsection{Codifferential, closedness, Laplacian and Hodge decomposition}
Consider the operator
$d^\star:\bigwedge^pM\to\bigwedge^{p-1}M$ defined as
\begin{equation}
d^\star\alpha^p\equiv -(-1)^{n(p+1)}\star d\star\alpha^p,
\end{equation}
which is called the co-differential. It is trivial to show that  $d^\star d^\star\alpha=0$. 
We can interpret it as the adjoint of $d$ relative to the inner product of forms since a simple calculation shows that
\begin{equation}
d(\alpha^{p-1}\wedge\star\beta^p)=d\alpha^{p-1}\wedge\star\beta-\alpha\wedge \star d^\star\beta^p.
\end{equation}

 A form $\alpha^p$ is said to be closed if $d\alpha^p=0$ everywhere on the manifold, and exact if $\alpha^p=d\beta^{(p-1)}$ for some $\beta^{(p-1)}$. Similarly, we speak of co-closed and co-exact forms.    
Exact forms are closed by Poincar\'e lemma. The converse is always true locally. Globally, it depends on the topology of the manifold. 

The Helmholtz decomposition theorem  (a freak of three-dimensional spaces) is subsumed into the much more general Hodge decomposition, which states that given a $p-$form $\alpha^p$ in a closed Euclidean  manifold, there are three  forms, $\beta^{p-1},\beta^p$ and $\beta^{p+1}$ such that 
\begin{equation}
\alpha^p=\beta^p+d^\star\beta^{p+1}+d\beta^{p-1},
\end{equation}
where $\beta^p$ is both closed and co-closed, that is an harmonic form. In other words, any form can be written as the sum of a harmonic form, plus an exact form plus a co-exact form. Moreover, the exact and coexact forms are unique (that is, the $d\beta^{p-1}$ is unique, not the $\beta^{p-1}$, and likewise for the the other term) and mutually orthogonal. Finally, it can be shown (Hodge's theorem, which is at the base of the decomposition) that 
$\beta^{p+1}=d\gamma^p$ and $\beta^{p-1}=d^\star\gamma^p$, so that 
if we introduce the operator $\nabla^2\equiv (dd^\star+d^\star d)$ (which generalizes the Laplacian to $p-$forms), 
\begin{equation}
\alpha^p-\beta^p=\nabla^2\gamma^p.
\end{equation}

\subsection{Pullbacks and change of coordinates}\label{App:Pullbacks}
Consider a $p-$form $\alpha^p\in\bigwedge^p\mathbb{M}^n$. Suppose we have a map $S:\mathbb{V}^q\to\mathbb{M}^n$. Note that $q$ does not need to be equal to $n$. This map can be used to pull back (hence the name) the form $\alpha^p$ to the manifold ${\mathbb{V}^q}$ via a simple composition rule. That is, let ${\cal A}^p$ be a $p-$dimensional surface in $\mathbb{V}^q$. Then $S({\cal A}^p)$ is a $p-$dimensional surface in $\mathbb{M}^n$, and we can feed it to $\alpha^p$ to get its value. We denote the process with 
\begin{equation}
S^*(\alpha^p)\in\bigwedge^p\mathbb{V}^q.
\end{equation}
Obviously, if $q<p$ then $S^*(\alpha^p)=0$.  A change of coordinates is a special pullback with $q=n$. The reason why exterior calculus provides a true coordinate-free description of geometry is given by the following relationships  which are stated without proof (here $\alpha^s$ and $\beta^l$ are arbitrary forms):
\begin{gather}
S^\star(d\alpha^s)=d(S^\star\alpha^s),\\
S^\star(\alpha^s\wedge\beta^l)=S^\star(\alpha^s)\wedge S^\star(\beta^l).\\
S^\star(\alpha^s+\beta^s)=S^\star(\alpha^s)+S^\star(\beta^s).
\end{gather}

Since the structure of the graded algebra depends on the dimension of the manifold, we cannot expect the Hodge star to be transparent to general pullbacks. However, when the pullback is between Euclidean manifolds with the same dimension, e.g., a change of coordinates, then we do have
\begin{equation}
S^*(\star\alpha^p)=\star(S^*\alpha^p).
\end{equation}
Likewise, there is no way to unambiguously transfer vectors between spaces of different dimension. When $S$ is a change of coordinates, then 
\begin{equation}
S^*(i_{\mathbf{v}}\alpha^p)=i_{\hat{\mathbf{v}}}(S^*\alpha^p),
\end{equation}
where $\hat{\mathbf{v}}$ is the pushforward of the vector (i.e., the vector in the new coordinates). 
Taken together, these results show that the machinery of exterior calculus is completely transparent to changes of coordinates, in the sense that a change of coordinates does not introduce "extra" terms.  

\section{From Lagrange to Lie}
\subsection{Lie Groups}\label{App:LieG}
The concept of group plays a fundamental role in any modern field theory. A group is a set $G$ with an associative binary operation $\cdot$ (a composition law) characterized by the following axioms
\begin{gather}
\mathbf{Closure}:\, \mbox{if}\,a,b\in G,\,a\cdot b\in G.\\
\mathbf{Existence\,\, of\,\, Identity}:\,\exists\, I\in G,\,\mbox{such that}\,\, \forall\,a\in G\,\,a\cdot I=I\cdot a=a.\\
\mathbf{Invertibility}:\,\forall\,a\in G,\,\exists\, a^{-1}\,\,\mbox{such that}\,a\cdot a^{-1}=a^{-1}\cdot a=I.
\end{gather}
A Lie group is a Group which is also a differentiable manifold. Of particular interest here are 1-dimensional (or 1-parameter) Lie groups whose members ($s$ being the parameter) $G(s):\mathbb{M}^n\to\mathbb{M}^n$ are maps of the manifold onto itself. 
We can always choose the parameter such that $G^{-1}(s)=G(-s)$. If $x^i(s)=G(s)^i(x^1,\ldots,x^n)$ is the "trajectory" of a point under the action of a group, we can consider 
\begin{equation}
\frac{\mathrm{d}x^j}{\mathrm{d}s}\left|_{s=0}\right.=v^i\frac{\partial x^j}{\partial x^i}={\mathbf v}x^j,\label{eq:B4}
\end{equation}
where 
\begin{equation}
v^i=\lim_{s\to 0}\frac{G(s)^i(x^1,\ldots,x^n)-G(0)^i(x^1,\ldots,x^n)}{s},
\end{equation}
are the \textit{infinitesimals} of the group and ${\mathbf v}$ is the \textit{group operator}. Knowledge of the operator is enough to recover the group via the formal solution of (\ref{eq:B4}) 
\begin{equation}
G(s)=e^{s{\mathbf{v}}},
\end{equation}
from which we have the formal definition of the group operator as
\begin{equation}
{\bf v}=G(-s)\frac{\mathrm{d}G}{\mathrm{d}s}.
\end{equation}
\subsection{The Lie derivative}
A $p-$form associates a value to a $p-$surface. Under the action of $G(s)$ (as defined in the preceding section), the form and the surface will change.  
The Lie derivative of a form $\alpha^p$ w.r.t. the flow $\mathbf{v}$, denoted as ${\cal L}_{\mathbf{v}}(\alpha^p)$, is the rate of change of $\int_{\Gamma^p}\alpha^p$ as the form $\alpha^p$ and the surface $\Gamma^p$ change under the action of the group, that is
\begin{equation}
\frac{\mathrm{d}}{\mathrm{d}s}\int_{\Gamma^p(s)}\alpha^p=\int_{\Gamma^p(s)}{\cal L}_{\mathbf{v}}(\alpha^p).
\end{equation}
Fortunately for us, Cartan himself showed how to express the Lie derivative of forms in terms of exterior calculus operators \citep[see, e.g., ref.][p. 138]{cantwell_introduction_2002}
\begin{equation}
{\cal L}_{\bf v}(\alpha^p)=i_{\bf v}d\alpha^p+di_{\bf v}\alpha^p,
\end{equation}
which is often (and rightly so!) called Cartan's magic formula. 
The Lie derivative enjoys some highly non-trivial properties which are nonetheless trivial to derive with the algebra of exterior calculus. They  become extremely useful when working out physical properties. We have
\begin{equation}
d{\cal L}_{\bf v}(\alpha^p)={\cal L}_{\bf v}(d\alpha^p)
\end{equation}
and 
\begin{equation}
i_{\bf v}{\cal L}_{\bf v}(\alpha^p)={\cal L}_{\bf v}(i_{\bf v}\alpha^p).
\end{equation}
For the last equality to hold, Lie derivative and interior product have to be on the same vector! Indeed, a more general expression is 
\begin{equation}
{\cal L}_{\bf v}(i_{\bf w}\alpha^p)-i_{\bf w}{\cal L}_{\bf v}(\alpha^p)=i_{[{\bf v},{\bf w}]}\alpha^p,\label{eq:Lie_Lieb}
\end{equation}
where 
\begin{equation}
[{\bf v},{\bf w}]\equiv \left(v^i\frac{\partial w^j}{\partial x^j}-w^i\frac{\partial v^j}{\partial x^i}\right)\bm{\frac{\partial}{\partial x^j}}
\end{equation}
is the commutator of the two vectors.
Also, the Lie derivative is linear both in ${\bf v}$ and $\alpha$. The Lie derivative satisfies the Leibniz product rule 
\begin{equation}
{\cal L}_{\mathbf{v}}(\alpha^p\w\alpha^q)={\cal L}_{\mathbf{v}}(\alpha^p)\w\alpha^q+\alpha^p\w{\cal L}_{\mathbf{v}}\alpha^q.
\end{equation}
Finally, applied to a $0-$form, the Lie derivative is simply the projection of the gradient of the scalar over the group infinitesimal, and so the Lie derivative is simply the rate of change of the scalar along the flow induced by the group, that is it coincides with the standard Lagrangian or material derivative.

\subsection{Concluding remarks}
This brief overview of the machinery of exterior calculus should give the reader a basic primer of the tools that were used in the main body of the paper. 
A few points are worth being emphasized:
\begin{enumerate}
\item The zoo  of three-dimensional vector calculus operators (Div, Grad, Curl) is subsumed into a single unifying operator $d$. 
\item The main structures and operators are coordinate-independent.
  In other words, exterior calculus, unlike tensor calculus,  de-emphasizes the local coordinate system and allows to focus on the physical entities, whereas  the latter relies heavily on the {\em natural frames} associated to the local coordinates. 
 \item It is important to emphasize that we have dealt {\em only} with scalar-valued forms, which are enough to build an elementary  theory of fluids. The reader who may wander if it is possible to consider vector (or more generally tensor) valued forms is referred to \citet{vargas_differential_2014} and \citet{Westenholz81} for a discussion of what further structures are needed.

\end{enumerate}

\bibliography{Zotero,Local_bib}
\bibliographystyle{abbrvnat} 
\end{document}